\documentclass[prb,twocolumn,showpacs]{revtex4}
\usepackage{graphicx}
\usepackage{soul}
\usepackage{color}
\begin{document}
\title{Chimeras in SQUID Metamaterials}
\author{N. Lazarides$^{1,2}$, G. Neofotistos$^{1}$, G. P. Tsironis$^{1,2,3}$}
\affiliation{
$^{1}$Crete Center for Quantum Complexity and Nanotechnology, 
      Department of Physics, University of Crete, P. O. Box 2208, 71003 Heraklion, 
      Greece; \\
$^{2}$Institute of Electronic Structure and Laser,
      Foundation for Research and Technology-Hellas, P.O. Box 1527, 71110 Heraklion, Greece; \\
$^{3}$Department of Physics, School of Science and Technology, Nazarbayev University,
      53 Kabanbay Batyr Ave., Astana 010000, Kazakhstan
}
\date{\today}
\begin{abstract}
Regular lattices comprising Superconducting QUantum Interference Devices (SQUIDs) form
magnetic metamaterials exhibiting extraordinary properties, including tuneability, 
dynamic multistability, and negative magnetic permeability. The SQUIDs in a metamaterial
interact through non-local, magnetic dipole-dipole forces, that makes it possible for
counter-intuitive dynamic states referred to as {\em chimera states} to appear; the latter
feature clusters of SQUIDs with synchronous dynamics which coexist with clusters exhibiting 
asynchronous behavior. The spontaneous appearence of {\em chimera states} is demonstrated
numerically for one-dimensional SQUID metamaterials driven by an alternating magnetic 
field in which the
fluxes threading the SQUID rings are randomly initialized; then, chimera states appear 
generically for sufficiently strong
initial excitations, which exhibit relatively long life-times. The synchronization and 
metastability levels of the chimera states are discussed in terms of appropriate measures.
Given that both one- and two-dimensional SQUID metamaterials
have been already fabricated and investigated in the lab, the presence of a chimera state 
could in principle be detected with presently available experimental set-ups.
\end{abstract}
\maketitle
\section{Introduction}  
\begin{figure}[!h]
\includegraphics[angle=0, width=0.85 \linewidth]{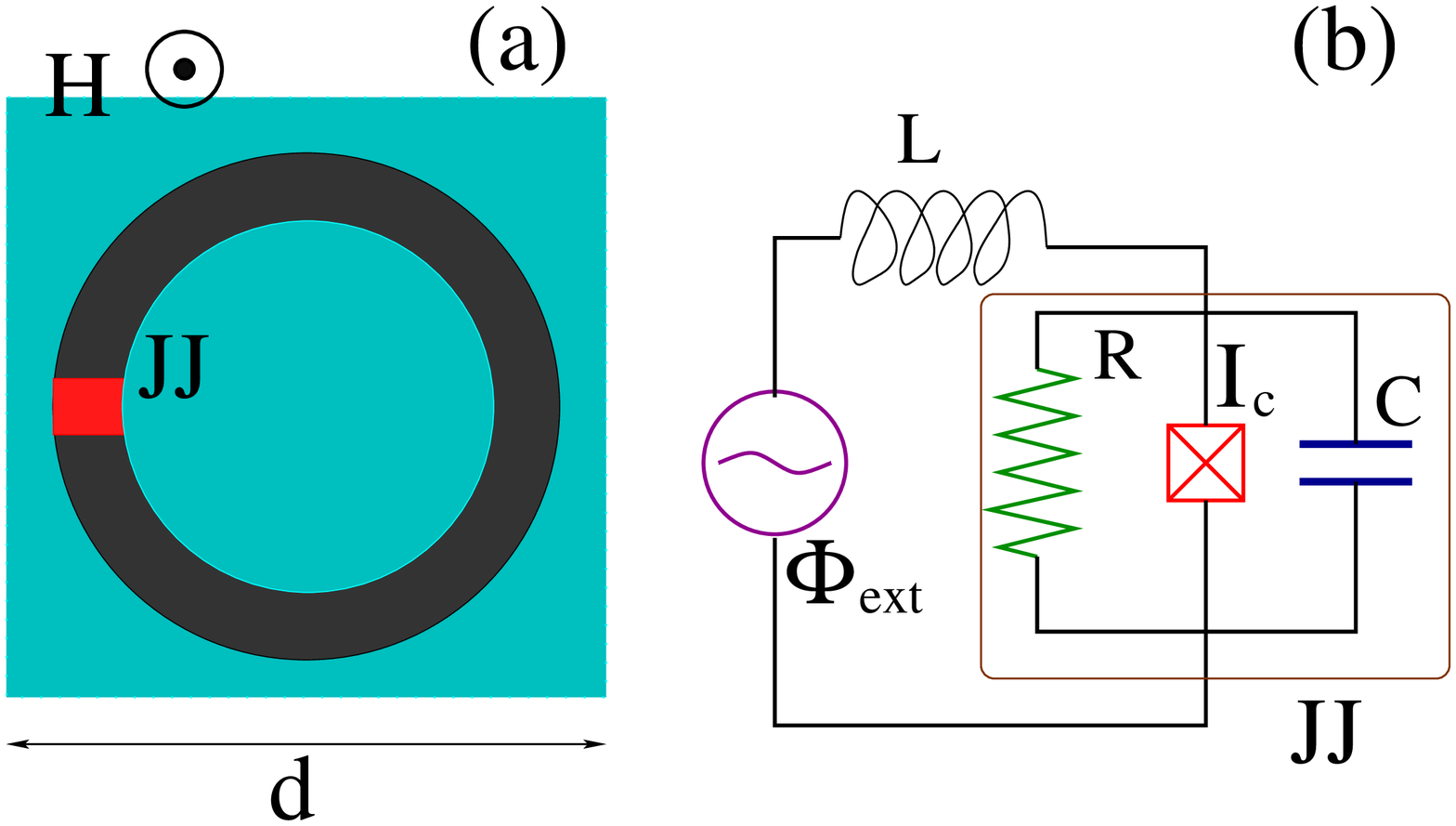} 
\includegraphics[angle=0, width=0.88 \linewidth]{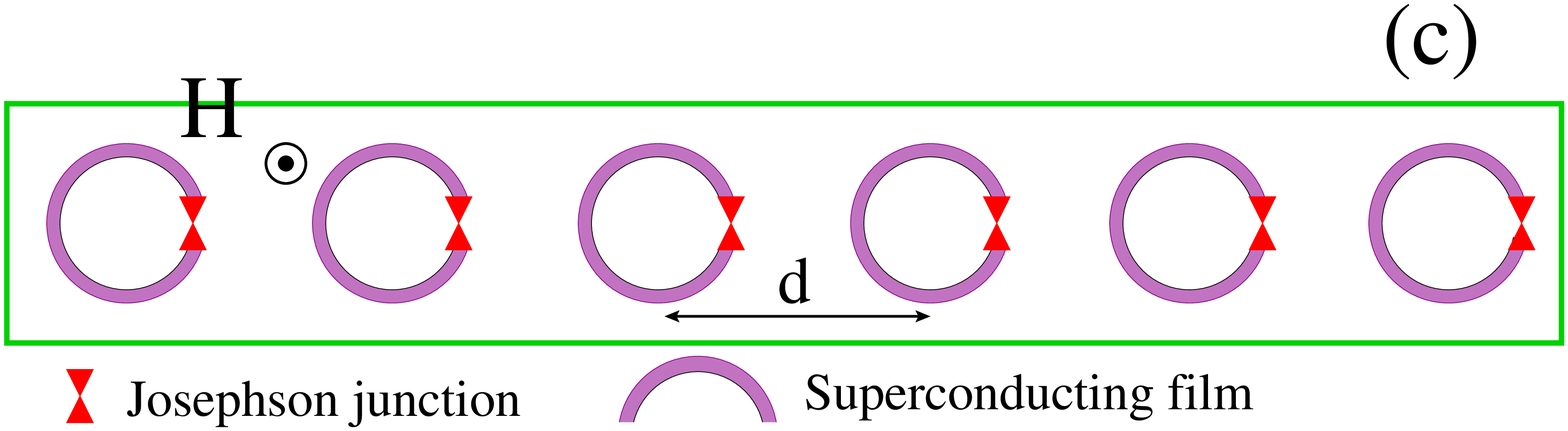} 
\caption{(Color online)
Schematic of an rf SQUID in an alternating magnetic field ${\bf H}(t)$ (a), and equivalent 
electrical circuit (b). The real Josephson junction is represented by the circuit elements
that are in the brown-dashed box.
(c) A one-dimensional SQUID metamaterial is formed by repetition of the squared unit cell of 
side $d$ shown in (a). 
\label{fig1}
}
\end{figure}

Superconducting QUantum Interference Device
(SQUID) metamaterials constitute a subclass of superconducting artificial media 
whose function relies both on the geometry and the extraordinary properties of 
superconductivity and the Josephson effect \cite{Anlage2011,Jung2014}. Recent experiments
on both one- and two-dimensional radio-frequency (rf) SQUID metamaterials in the superconducting 
state have 
demonstrated their wide-band tuneability, significantly reduced losses, and dynamic 
multistability \cite{Jung2013,Butz2013a,Trepanier2013,Jung2014b,Jung2014}.
The simpest version of an rf SQUID consists of
a superconducting ring interrupted by a Josephson junction \cite{Josephson1962} 
(Fig. \ref{fig1}a); 
that device is a highly nonlinear resonator with a 
strong response to applied magnetic fields.
SQUID metamaterials exhibit peculiar magnetic properties including negative diamagnetic 
permeability that were predicted both for the quantum \cite{Du2006} and the classical 
\cite{Lazarides2007} regime.
The applied alternating fields induce (super)currents in the SQUID rings, coupling them
together through dipole-dipole magnetic forces; although weak due to its magnetic nature, 
that interaction couples the SQUIDs non-locally since it falls-off as the inverse cube of
their center-to-center distance.

The study of networks of interacting nonlinear elements pervades all of science,
from neurobiology to statistical physics, often revealing remarkable aspects of
collective behaviour \cite{Strogatz2001}. The effect of non-local interactions,
which constitutes the "dark corner" of nonlinear dynamics, has been extensively 
investigated in the last decade and has unveiled collective dynamic effects such 
as synchronization \cite{Strogatz2000,Acebron2005}, pattern formation 
\cite{Battogtokh1999}, and Turing instabilities \cite{Viana2011}.
Recently, a state with a counter-intuitive structure, usually referred to as a 
{\em "chimera state"}, was discovered in numerical simulations of non-locally 
coupled oscillator arrays \cite{Kuramoto2002}. Since then, an intense theoretical
\cite{Abrams2004,Kuramoto2006,Omelchenko2008,Abrams2008,Pikovsky2008b,Ott2009,Martens2010,Omelchenko2011,Yao2013,Omelchenko2013,Hizanidis2014,Zakharova2014,Bountis2014,Yeldesbay2014} 
and experimental
\cite{Tinsley2012,Hagerstrom2012,Wickra2013,Nkomo2013,Martens2013,Schonleber2014,Viktorov2014,Rosin2014,Schmidt2014b,Gambuzza2014,Kapitaniak2014} 
activity has been initiated.  
A {\em "chimera state"} is characterized by the coexistence of synchronous and 
asynchronous clusters (subgroups) of oscillators, even though they are coupled 
symmetrically and they are identical \cite{Smart2012,Panaggio2014}.
In the present work, it is demonstrated numerically that this remarkable collective 
dynamic behaviour emerges in SQUID metamaterials which are driven by an alternating magnetic 
field in the presence of weak dissipation.
The SQUID metamaterial model which has been used previously in the weak and local coupling 
regime, in which each SQUID is only coupled to its nearest neighbors, for the  investigation 
of intrinsic localization and tuneability effects 
\cite{Lazarides2008a,Lazarides2012,Lazarides2013b}, is being extended to account for 
non-local magnetic interactions.

Chimera states with very long life-times are obtained by initializing randomly the fluxes
threading the SQUID rings; the subsequent dynamic evolution yields spontaneous formation 
of clusters
of SQUIDs that exhibit synchronous and asynchronous flux oscillations for sufficiently strong 
initial flux excitations. Their existence requires multistability of the individual SQUIDs
at frequencies in the region around that of the driving field. 
Individual SQUID states with high current amplitude may customarily be reached from basins
of attraction which are much smaller than those of the states with low current amplitude.
The number of possible states in a SQUID metamaterial is not merely the sum
of the combinations of individual SQUID states, since their collective behavior may provide 
many more possibilities. Chimera states are such an example, where qualitatively distinct
and counter-intuitive states emerge from collective dynamics. Although chimera states 
are generally regarded to be metastable \cite{Shanahan2010,Wildie2012}, or even chaotic
transients \cite{Wolfrum2011}, there are also examples where they are at the global minimum
of a system in thermal equilibrium, as has been predicted for Ising spins \cite{Singh2011}.
The level of synchronization and metastability of the chimera states can be however quantified 
by several measures \cite{Shanahan2010,Wildie2012}. Many different types of non-local 
interactions have been employed in the literature, often exponentially decaying, that allow 
a particular system to reach a chimera state. Importantly, chimera states in globally 
coupled systems that have been recently demonstrated \cite{Yeldesbay2014,Sethia2014}, may
lead to a revision of their existence criteria. 
The magnetic dipole-dipole interaction between SQUIDs, considered here,  although it is 
still short-ranged \cite{Campa2009}, it falls-off much slower than the exponentially 
decaying one and it is capable of rendering the system able to support chimera states.

In the next Section, the non-locally coupled SQUID metamaterial model is presented, and 
the linear frequency dispersion is obtained. In the same Section, the complex synchronization
(Kuramoto-type) parameter is defined along with a measure of metastability.
In Section III, the numerically obtained spatiotemporal evolution of several chimera states 
are shown, and compared with those of the corresponding locally coupled SQUID metamaterial.
The magnitude of the synchronization parameter is monitored in time both for the non-locally 
and the locally coupled SQUID metamaterials, and the differences are indicated.
The metastability of the chimera states is discussed in terms of the full-width half-maximum
of the distribution of the magnitude of the synchronization parameter at each time-instant.
The conclusions are given in the last Section.
\begin{figure}[!h]
\includegraphics[angle=0, width=0.95 \linewidth]{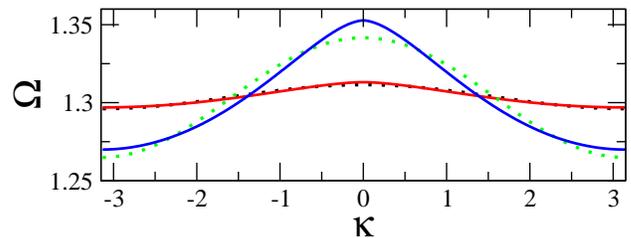} 
\caption{(Color online)
Frequency dispersion of the SQUID metamaterial with non-local coupling, for 
$\beta = 0.1114$ ($\beta_L \simeq 0.7$)
and $\lambda_0=-0.05$ (blue curve); $-0.01$ (red curve).
The corresponding dispersions for nearest neighbor coupling are shown as green and black
dotted curves, respectively. 
\label{fig2}
}
\end{figure}

\section{SQUID Metamaterial Modelling and Measure of Synchronization}
Consider a one-dimensional linear array of $N$ identical SQUIDs coupled together 
magnetically through dipole-dipole forces. The magnetic flux $\Phi_n$ threading the $n-$th
SQUID loop is
\begin{eqnarray}
\label{01}
  \Phi_n =\Phi_{ext} +L\, I_n +L\, \sum_{m\neq n} \lambda_{|m-n|} I_m ,
\end{eqnarray}
where the indices $n$ and $m$ run from $1$ to $N$, 
$\Phi_{ext}$ is the external flux in each SQUID, 
$\lambda_{|m-n|} =M_{|m-n|}/L$ is the dimensionless coupling coefficient between the SQUIDs 
at positions $m$ and $n$, with $M_{|m-n|}$ being their corresponding mutual inductance,
and
\begin{eqnarray}
\label{02}
    -I_n =C\frac{d^2\Phi_n}{dt^2} +\frac{1}{R} \frac{d\Phi_n}{dt} 
                                +I_c\, \sin\left(2\pi\frac{\Phi_n}{\Phi_0}\right) 
\end{eqnarray}
is the current in each SQUID given by the resistively and capacitively shunted junction 
(RCSJ) model \cite{Likharev1986}, with  $\Phi_0$ and $I_c$ being the flux quantum and the 
critical current of the Josephson junctions, respectively.  Within the RCSJ framework,
$R$, $C$, and $L$ are the resistance, capacitance, and self-inductance of the SQUIDs'
equivalent circuit (Fig. \ref{fig1}b). Combination of Eqs. (\ref{01}) and (\ref{02}) gives
\begin{eqnarray}
\label{05}
  C\frac{d^2\Phi_n}{dt^2} +\frac{1}{R} \frac{d\Phi_n}{dt}
    +\frac{1}{L} \sum_{m=1}^N  \left( {\bf \hat{\Lambda}}^{-1} \right)_{nm} 
         \left( \Phi_m -\Phi_{ext} \right) 
   \nonumber \\
   +I_c\, \sin\left(2\pi\frac{\Phi_n}{\Phi_0}\right) =0 ,
\end{eqnarray}
where ${\bf \hat{\Lambda}}^{-1}$ is the inverse of the $N\times N$ coupling matrix 
\begin{eqnarray}
\label{04}
  {\bf \hat{\Lambda}} = \left\{ \begin{array}{ll}
        1, & \mbox{if $m= n$};\\
        \lambda_{|m-n|} =\lambda_0 \, |m-n|^{-3}, & \mbox{if $m\neq n$},\end{array} \right.   
\end{eqnarray}
with $\lambda_0$ being the coupling coefficient betwen nearest neighboring SQUIDs.
\begin{figure*}[!t]
\includegraphics[angle=0, width=0.95 \linewidth]{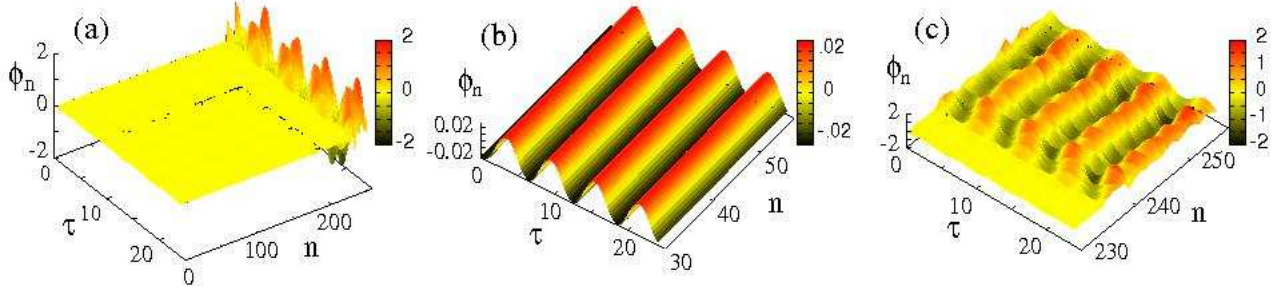} 
\caption{(Color online)
Spatio-temporal evolution of the normalized fluxes $\phi_n$ threading the SQUID rings 
during four driving periods $T=5.9$ for $N=256$, $\gamma=0.0022$, $\lambda_0=-0.05$,
$\beta=0.1114$ ($\beta_L\simeq 0.7$), $\phi_{ac}=0.015$, and $\phi_R =0.85$.
(a) for the whole SQUID metamaterial; 
(b) for part of the metamaterial that belongs to the coherent cluster;
(c) for part of the metamaterial that includes the incoherent cluster.
\label{fig3}
}
\end{figure*}
In normalized form Eq. (\ref{05}) reads ($n=1,...,N$)
\begin{eqnarray}
\label{06}
  \ddot{\phi}_n +\gamma \dot{\phi}_n +\beta \sin\left( 2\pi \phi_n \right) 
    =\sum_{m=1}^N \left( {\bf \hat{\Lambda}}^{-1} \right)_{nm} 
         \left( \phi_{ext} -\phi_m \right),
\end{eqnarray}
where the frequency and time are normalized to $\omega_0 =1/\sqrt{LC}$ and its inverse 
$\omega_0^{-1}$, respectively, the fluxes and currents are normalized to $\Phi_0$
and $I_c$, respectively, 
the overdots denote derivation with respect to the normalized temporal variable, $\tau$,
$\phi_{ext} =\phi_{ac} \cos(\Omega \tau )$, with $\Omega=\omega/\omega_0$ being 
the normalized driving frequency, and 
\begin{equation}
\label{14}
   \beta=\frac{I_c L}{\Phi_0} =\frac{\beta_L}{2\pi}, \qquad
   \gamma=\frac{1}{R} \sqrt{ \frac{L}{C} } 
\end{equation}
is the SQUID parameter and loss coefficient, respectively.
The value of $\beta_L$ determines whether a SQUID is hysteretic or non-hysteretic
($\beta_L >1$ and $\beta_L <1$, respectively).

Linearization of Eq. (\ref{06}) around zero with $\gamma=0$ and $\phi_{ext} =0$ gives for 
the infinite system
\begin{eqnarray}
\label{07}
  \ddot{\phi}_n +\left[ \beta_L +\left( {\bf \hat{\Lambda}}^{-1} \right)_{nn} \right] \phi_n
    +\sum_{m\neq n} \left( {\bf \hat{\Lambda}}^{-1} \right)_{nm} \phi_m =0 . 
\end{eqnarray}
By substitution of $\phi_n=A\, \exp[i(\kappa n -\Omega \tau)]$ into Eq. (\ref{07}), with 
$\kappa$ being the wavevector normalized to $d^{-1}$, and using
\begin{eqnarray}
\label{08} 
  \sum_{m\neq n} \left( {\bf \hat{\Lambda}}^{-1} \right)_{nm} e^{i\kappa(m-n)} 
       =2 \sum_{m=1}^\infty \left( {\bf \hat{\Lambda}}^{-1} \right)_{m} \cos(\kappa m) ,
\end{eqnarray}
where $m$ is the "distance" from the main diagonal of ${\bf \hat{\Lambda}}^{-1}$, we get
\begin{eqnarray}
\label{09} 
   \Omega =\sqrt{ \Omega_{0}^2 
     +2 \sum_{m=1}^\infty \left( {\bf \hat{\Lambda}}^{-1} \right)_{m} \cos(\kappa m) } ,
\end{eqnarray}
where $\Omega_{0}^2 =\beta_L +\left( {\bf \hat{\Lambda}}^{-1} \right)_{nn} \simeq \beta_L +1$.
Note that for the infinite system the diagonal elements of the inverse of the coupling 
matrix $\left( {\bf \hat{\Lambda}}^{-1} \right)_{nn}$ have practivally the same value 
which is slightly larger than unity.
The frequency $\Omega_{0}$ is very close to the resonance frequency of individual SQUIDs,
$\Omega_{SQ} =\sqrt{\beta_L +1}$.
The nonlocal frequency dispersion Eq. (\ref{09}) is slightly different from that obtained 
with only nearest-neighbor coupling (Fig. \ref{fig2}).

Eqs. (\ref{06}) are numerically integrated in time using a fourth-order Runge-Kutta algorithm 
with fixed time stepping, typically $\Delta t=0.02$, with $\phi_n (\tau=0)$ randomly chosen 
from a flat, zero mean distribution in $[-\phi_R/2, +\phi_R/2]$ and $\dot{\phi}_n (\tau=0)=0$
for all $n$. 
The boundary conditions
\begin{equation}
\label{11} 
   \phi_0 (\tau) =0 , \qquad \phi_{N+1} (\tau) =0 ,
\end{equation}
are used to account for the termination of the structrure in a finite system. 
The degree of synchronization for a given cluster of SQUIDs or for the whole SQUID metamaterial 
having $M$ elements is quantified by introducing a complex, Kuramoto-type
synchronization parameter $\Psi(\tau)$, defined as
\begin{equation}
\label{10} 
  \Psi(\tau) = \frac{1}{M} \sum_{m=1}^M e^{i [2\pi \phi_m (\tau)]} .
\end{equation}
The magnitude of the synchronization parameter, $|\Psi(\tau)|$, provides a global 
(for the whole metamaterial) or local (within a cluster) measure of spatial coherence at
time-instant $\tau$. The value of $|\Psi(\tau)|$ lies in the interval $[0,1]$,
where the values $0$ and $1$ correspond to complete desynchronization and synchronization,
respectively. The mean synchrony level $\bar{|\Psi|}$, which is an index of the global 
synchronization level, is defined as the average of $|\Psi(\tau)|$ over the time interval 
of integration \cite{Wildie2012}, while the variance of $|\Psi(\tau)|$, 
$\sigma_{|\Psi|}^2$, captures how the degree of synchrony fluctuates in time.
Fluctuations of the degree of synchronization have been associated with the existence of
metastable states and therefore $\sigma_{|\Psi|}^2$ is indicative of the system's metastability
level \cite{Shanahan2010,Wildie2012}.

The parameters used in the simulations are rather typical for SQUID metamaterials. 
In Ref. \cite{Trepanier2013}, the values of the SQUIDs' inductance, resistance,
and capacitance are $L=0.12~nH$, $R=840~\Omega$, and $C=0.65~pF$, respectively,
while the critical 
current is $3.7~\mu A$ and $1.2~\mu A$ at temperature $T=4.2~K$ and $6.5~K$, respectively.
Inserting these values in the first of Eqs. (\ref{14}), we get for $\beta_L$ the values 
$0.44$ and $1.3$ 
at $T=4.2~K$ and $6.5~K$, respectively. Here we use $\beta_L =0.7$, which lies in between
those values, while our value of the amplitude of the alternating field $\phi_{ac}=0.015$ 
lies within 
the range of values used in the experiments ($\phi_{ac} \simeq 0.006 - 0.05$). 
The coupling coefficient between nearest neighbors in Ref. \cite{Trepanier2013} 
has been estimated to be $\lambda_0 \simeq-0.02$, using a simple approximation scheme in 
which the SQUIDs
are regarded as thin rings. However, a large part of the area of the actual SQUID metamaterial 
is filled by superconducting material that the field cannot penetrate; it is thus expected 
that more magnetic flux than that predicted by the simple approximation will find its way 
through the SQUID rings, increasing thus considerably $\lambda_0$ (we use $\lambda_0=-0.05$). 
The value of the loss coefficient in our simulations is $\gamma \simeq 0.002$, while 
from the second of Eqs. (\ref{14}) and the values given in Ref. \cite{Trepanier2013} we get
$\sim 0.02$. Although the losses in the SQUIDs can be reduced considerably by lowering the
temperature without affecting much the critical currents of the junctions, we have checked 
that chimera states also exist for $\gamma$ of the order of $\sim 0.02$ when compensated
by a larger ac field amplitude $\phi_{ac}$.

Note that the comparison of the parameters used in the simulations of the one-dimensional
SQUID metamaterial are compared with those in Ref. \cite{Trepanier2013}, in which
the experiments have been performed on two-dimensional arrays, merely to show that these 
are realistic. In this work, we do not attempt to simulate a particular SQUID metamaterial.
Moreover, the dimensionality of the system does not affect significantly the estimation
of the parameters that are necessary for the simulations, since they can be estimated 
either by the individual SQUID properties (i.e., $\beta$, $\gamma$) or by a pair of 
SQUIDs (i.e., $\lambda_0$).

\section{Chimera States: Levels of Synchronization and Metastability}
Chimera states are very sensitive to slight changes of the model parameters, the parameters
of the driving field, as well as the integration parameters, e.g., the value of the time-step,
and there is indeed a large number of such states available in a SQUID metamaterial.
The chosen time-step $\Delta \tau=0.02$, in particular, is a typical choice for systems of
nonlinear oscillators that gives reliable results. Decreasing of the time-step (i.e., to 
$\Delta \tau=0.01$) leads the SQUID metamaterial to a different chimera state due to their 
high metastability; that state may be either more or less synchronized than the previous one,
depending on the other parameters. For the chosen parameters, the system reaches spontaneously 
dynamical states where synchronous (coherent) and asynchronous (incoherent) clusters of 
SQUIDs coexist, for most of the initial flux configurations with $\phi_R \sim \Phi_0$.
A typical spatiotemporal flux pattern obtained after $10^7$ time units (t.u.) of integration
is shown in Fig. \ref{fig3}a, where the evolution of the $\phi_n$s is monitored during four 
driving periods $T=2\pi/\Omega$. Two different domains of the array can be distinguished, 
in which the fluxes through the SQUIDs are oscillating either with low or high amplitude.
The enlargement of two particular subdomains shown in Figs. \ref{fig3}b and \ref{fig3}c
reveals the main feature of a chimera state; besides the difference in the oscillation 
amplitudes (i.e., low-high), there are different dynamic behaviors: the low-amplitude 
oscillations are completely synchronized (Fig. \ref{fig3}b) while the high-amplitude ones are 
desynchronized both in phase and amplitude (Fig. \ref{fig3}c).
Note that since the SQUID metamaterial is driven at a particular frequency $\Omega$, there
can be no net frequency drift as in phase oscillators \cite{Kuramoto2002}; instead, the 
period of each SQUID in the asynchronous cluster fluctuates around that of the driver, $T$. 

\begin{figure}[!h]
\includegraphics[angle=0, width=0.95 \linewidth]{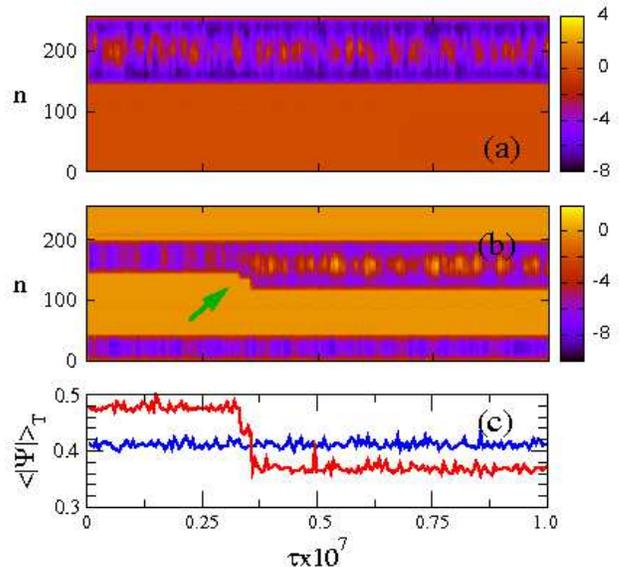} 
\caption{(Color online)
Density plot of the fluxes $\phi_n$ on the $n - \tau$ plane for a non-locally coupled
SQUID metamatgerial with $T=5.9$, $N=256$, 
$\gamma=0.0021$, $\lambda_0=-0.05$, $\beta = 0.1114$ ($\beta_L\simeq 0.7$), $\phi_{ac}=0.015$,
and $\phi_R =0.9$ (a); $\phi_R =0.8$ (b). The green arrow indicates sudden expansions of the 
corresponding asynchronous cluster. 
(c) The corresponding magnitude of the Kuramoto parameter averaged over the driving period $T$,
$<|\Psi (\tau)|>_T$ as a function of time $\tau$;
the blue and red curves are obtained for the chimera state shown in (a) and (b), respectively.
\label{fig4}
}
\end{figure}

In Figs. \ref{fig4}a and  \ref{fig4}b, the long-term spatiotemporal evolution for the fluxes 
$\phi_n$ is shown as a density plot on the $n-\tau$ plane for two different initial flux
configurations; the values of the $\phi_n$s are obtained at time-instants that are multiples
of the driving period $T$, so that uniform (non-uniform) colorization indicates synchronous
(asynchronous) dynamics.
In Fig. \ref{fig4}a, two large clusters of SQUIDs have been apparently formed spontaneously, 
one with coherent and the other with incoherent dynamics. More SQUID clusters (two with 
synchronous and two with asynchronous behaviour) can be observed in Fig. \ref{fig4}b, in 
which the effect of metastability is reflected in the sudden expansions of the upper 
asynchronous cluster at around $\tau \sim 0.35\times 10^7$ t.u. (green arrow). In the 
corresponding time-dependent magnitudes of the synchronization parameter averaged over the 
driving period $T$, $<|\Psi (\tau)|>_T$, those sudden expansions make themselves apparent 
as abrupt changes towards lower synchronization levels (Fig. \ref{fig4}c).
\begin{figure}[!t]
\includegraphics[angle=0, width=0.95 \linewidth]{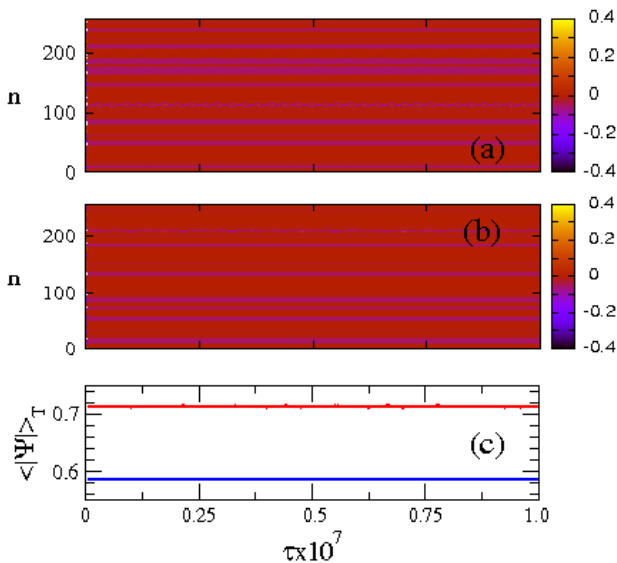} 
\caption{(Color online)
Density plot of the fluxes $\phi_n$ on the $n - \tau$ plane for a locally coupled SQUID 
metamaterial with $T=5.9$, $N=256$, 
$\gamma=0.0021$, $\lambda_0=-0.05$, $\beta_L=0.7$ ($\beta \simeq 0.1114$), $\phi_{ac}=0.015$,
and $\phi_R =0.9$ (a); $\phi_R =0.8$ (b). 
(c) The corresponding magnitude of the Kuramoto parameter averaged over the driving period $T$,
$<|\Psi (\tau)|>_T$ as a function of time $\tau$;
the blue and red curves are obtained for the non-uniform states shown in (a) and (b), respectively.
\label{fig5}
}
\end{figure}
By changing the coupling between the SQUIDs from non-local to local,
i.e., by taking into account the coupling of each SQUID only to its nearest-neighbours,
we obtain the corresponding patterns and  $<|\Psi (\tau)|>_T$s shown in Fig. \ref{fig5}.
The differences in the patterns of Figs. \ref{fig4} and \ref{fig5} are merely due to the 
change of the coupling from non-local to local.
As can be observed in Figs. \ref{fig5}a and \ref{fig5}b,
the SQUID metamaterial states are not uniform since a number of clusters have been formed;
however, the dynamics within all these clusters is synchronous. The different states occupied 
by the SQUIDs in the metamaterial are close to one of the available stable states of individual
SQUIDs. Although the dynamics within each
cluster is synchronized, the clusters are not synchronized to each other, resulting in a low
degree of synchronization (Fig. \ref{fig5}c). The comparison of Figs.  \ref{fig4}c and 
\ref{fig5}c is very illuminating: the curves obtained for non-locally coupled SQUIDs exhibit 
a much lower degree of synchronization than the corresponding ones for locally coupled SQUIDs
(e.g. for the blue curves, $<|\Psi (\tau)|>_T \simeq 0.42$ and $0.58$ for non-local and local
coupling, respectively). Moreover, the fluctuations in time are much stronger in the 
former case, due to the existence of asynchronous cluster(s).
For zero initial conditions, both the non-locally and locally coupled SQUID metamaterials
give uniform, competely synchronized states with $<|\Psi (\tau)|>_T$ practically equal to
unity (not shown).
Note that the fluxes in the asynchronous clusters in Figs. \ref{fig4}a and \ref{fig4}b have 
in general very high amplitude.
Similarly to the local coupling case, they correspond to high-flux-amplitude individual 
SQUID states which cannot be stabilized and therefore they cannot be reached by the SQUIDs
in the locally coupled SQUID metamaterial which occupy low-flux-amplitude states
(Figs. \ref{fig5}a and \ref{fig5}b). In the non-locally coupled SQUID metamaterial, 
to the contrary, the additional coupling provided by the second-nearest, third-nearest, 
etc., neighbors, is capable of keeping the SQUIDs close to these high-flux-amplitude
states for long time-intervals. 
\begin{figure}[!t]
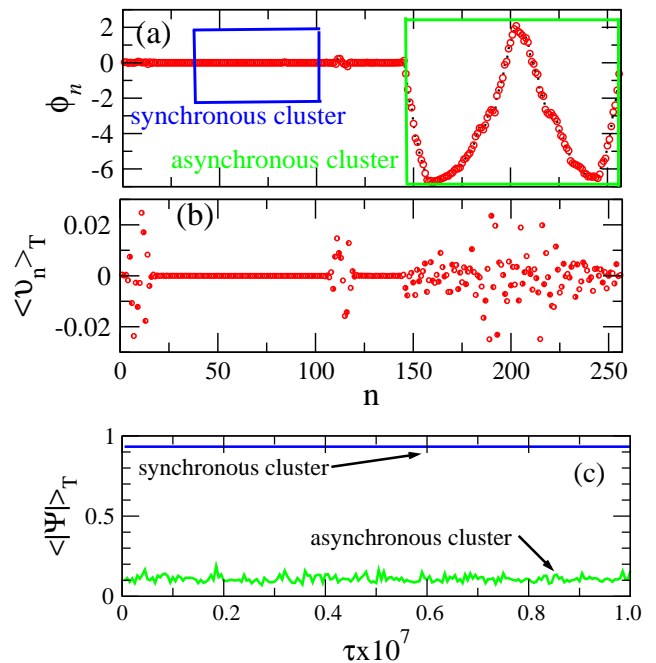

\includegraphics[angle=0, width=0.95 \linewidth]{Chimera-SQUID-Fig06ab.eps} \vspace{2mm} \\ 
\includegraphics[angle=0, width=0.95 \linewidth]{Chimera-SQUID-Fig06c.eps} \hspace{-8mm} 
\caption{(Color online)
(a) Spatial profile of the fluxes $\phi_n$ threading the SQUID rings at $\tau=10^7$ time 
    units for the parameters of Fig. \ref{fig4}. 
(b) The corresponding averaged voltage profile $<v_n (\tau)>_T =<\dot{\phi}_n (\tau)>_T$. 
(c) The magnitude of the synchronization parameter averaged over $T$, $<|\Psi|>_T$, as a 
    function of time
    $\tau$, calculated for the coherent cluster in the small-blue box (blue curve) and for 
    the incoherent cluster in the large-green box (green curve) in (a).
\label{fig6}
}
\end{figure}

The spatial profiles of $\phi_n$ and the time-derivatives of the fluxes averaged over $T$,
$<\dot{\phi}_n (\tau)>_T \equiv <v_n (\tau)>_T$, at the end of the integration time 
($\sim 10^7$ t.u.) of Fig. \ref{fig4}a are shown in Figs. \ref{fig6}a and \ref{fig6}b, 
respectively. Note that $v_n (\tau)$ is the instantaneous voltage across the Josephson
junction of the $n-$th rf SQUID. 
The $<v_n (\tau)>_T$s, corresponding to the time-derivatives of the phases of
the oscillators in Kuramoto-type phase oscillator models, are symmetrically distributed 
around zero. Notably, the shape of their distribution over the simulation period (not shown)
deviates significantly from a Gaussian profile due to correlations between different
non-locally interacting clusters.  
That type of pattern (Fig. \ref{fig6}b) is distinctly different than the standard one for 
chimera states in phase oscillator models \cite{Kuramoto2002}, while it resembles 
the corresponding one for a globally coupled system of complex Ginzburg-Landau oscillators
\cite{Sethia2014}. In Figs. \ref{fig6}a and \ref{fig6}b the synchronous clusters are indicated
by horizontal segments; we observe that besides the large incoherent cluster
extending from $n=143$ to $256$, there are actually two small ones (at around $n\sim 5$
and $n\sim 112$, more clearly seen in Fig. \ref{fig6}b) which are not visible in Fig. 
\ref{fig4}a.   
The calculated value of $<|\Psi|>_T$ as a function of $\tau$ for part of the coherent
cluster in the blue (small) box of Fig. \ref{fig6}a, that extends from $n=36$ to $100$,
is close to unity for all times (blue curve in Fig. \ref{fig6}c). The corresponding curve for 
the large incoherent cluster in the green (large) box has a significantly lower average 
value and strong fluctuations which do not decrease with time.

In a real SQUID metamaterial, there is naturally a spread in the parameter values of
individual SQUIDs due to fabrication-induced imperfections. The most important source of 
disorder comes from the critical current of the Josephson junctions, which are very 
sensitive to the thickness and quality of the insulating layer. As a result, disorder 
is induced in the SQUID parameters $\beta$ and their eigenfrequencies $\Omega_{SQ}$.
However, with the present technology the spread in $\beta$ could be less than $1\%$
of its nominal value, which makes the SQUID metamaterial weakly disordered, that could 
affect the observed chimera states. In order to check the robustness of the chimera states 
under that perturbation, we performed additional simulations in which $\beta$ is allowed 
to vary randomly in an interval up to $5\%$ of its nominal value. For random initial
flux configurations, we still obtain chimera states, although with different profiles,
due to their high metastability, while for zero initial fluxes we obtain low-amplitude
synchronized solutions. We thus conclude that chimera states are robust to weak
disorder, and they cannot be generated by weak disorder alone in SQUID metamaterials 
in the absence of substantial initial fluxes.

\begin{figure}[!h]
\includegraphics[angle=0, width=0.95 \linewidth]{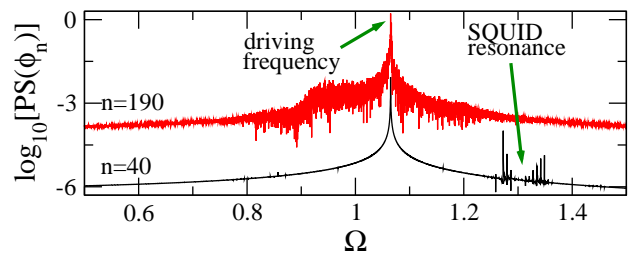} 
\caption{(Color online)
The {\em power spectra} of $\phi_n (\tau)$ in logarithmic scale for the SQUIDs with $n=40$
and $n=190$ that belong to the coherent (black curve) and the incoherent (red curve) cluster, 
respectively, of Fig. \ref{fig4}a. The arrow at right points at the eigenfrequency of
individual SQUIDs, $\Omega_{SQ} \simeq 1.3$. 
\label{fig7}
}
\end{figure}

The difference in the dynamic behaviour between SQUIDs in the coherent and incoherent clusters
is also reflected in the power spectra of $\phi_n (\tau)$. In Fig. \ref{fig7}, two such spectra,
one for a SQUID in the coherent cluster ($n=40$) and one in the incoherent cluster ($n=190$)
of Fig. \ref{fig4}a are shown in logarithmic scale for a range of frequencies around the 
fundamental 
(i.e., the driving) one. Note that for the chosen parameters, the resonance frequency
$\Omega_{SQ} \simeq \sqrt{ 1+\beta_L }$, of individual SQUIDs is at $\Omega_{SQ} \simeq 1.3$, 
while the linear band
of the SQUID metamaterial extends from $\Omega_{min} \simeq 1.27$ to 
$\Omega_{max} \simeq 1.35$ (extracted from the blue curve of Fig. \ref{fig2}). The driving 
frequency
is $\Omega \simeq 1.06$, well below the lower bound of the linear spectrum, $\Omega_{min}$. 
The spectrum for the SQUID at $n=40$ (black curve) exhibits very low noise levels and 
a strong peak at the driving frequency $\Omega$. The smaller peaks in the spectrum of the 
$n=40$ SQUID are also part of it, and they are located at frequencies within the linear band
of the SQUID metamaterial, i.e., in the range $[\Omega_{min}, \Omega_{max}]$. The longer 
arrow at right points at the resonance frequency (the eigenfrequency) of individual SQUIDs
which is located at $\Omega_{SQ} \simeq 1.3$. Note that only a small number of the 
eigenfrequencies of the SQUID metamaterial are excited in that spectrum, which seem to be 
selected by random processes.
To the contrary, the spectrum for the SQUID at $n=190$ exhibits significant fluctuations, 
the peak at the driving frequency, and in addition a frequency region around
$\Omega \sim 0.9-1.05$ where the average fluctuation level remains approximatelly constant, 
forming a shoulder that often appears in such spectra for SQUIDs in an asynchronous cluster
of a chimera state.

\begin{figure}[!t]
\includegraphics[angle=0, width=0.95 \linewidth]{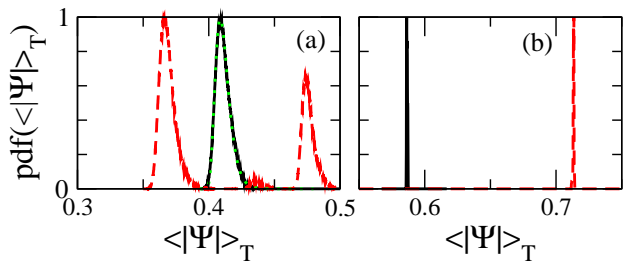} 
\caption{(Color online)
The distributions (divided by their maximum value) of $<|\Psi (\tau)|>_T$s at all instants
of the simulation period ($\sim 10^7$ time units with time-step $\Delta t=0.02$) for the 
states shown in Figs. \ref{fig4} and \ref{fig5}.
(a) for the chimera states of the non-locally coupled SQUID metamaterial shown in 
Fig. \ref{fig4}a (black-solid curve) and Fig. \ref{fig4}b (red-dashed curve).
The green-dotted curve is a fit with Eq. (\ref{12}).
(b) for the non-uniform states of the locally coupled SQUID metamaterial shown in 
Fig. \ref{fig5}a (black-solid curve) and Fig. \ref{fig5}b (red-dashed curve).
\label{fig8}
}
\end{figure}

In order to determine the metastability levels of the states obtained in Figs. \ref{fig4}
and \ref{fig5}, the distributions of $x \equiv <|\Psi (\tau)|>_T$s, $pdf(x)$, at all 
time-steps over the simulation period were calculated (Fig. \ref{fig8}). A transient period
of $100 T$ ($\sim 5900$ t.u.) was allowed, for which the data were discarded. 
The huge difference of the dispersion of the $x \equiv <|\Psi (\tau)|>_T$ values around 
their mean
$\bar{|\Psi|}$ between non-locally and locally coupled SQUID metamaterial shown in Figs. 
\ref{fig8}a and \ref{fig8}b, respectively, is readily observed. Consider first the black-solid
curves in these figures, which are actually not symmetric but they could fit satisfactorily
by an empirical skewed Gaussian function of the form \cite{Fraser1969,Rusch1973} 
\begin{equation}
\label{12} 
   pdf(x) =pdf_m \exp\left\{-\ln(2) 
       \left[ \frac{1}{b} \ln \left( 1 +\frac{2 b (x-x_m)}{D} \right) \right]^2 
        \right\} , 
\end{equation}
where $pdf_m =pdf(x_m)$ is the maximum of the distribution, 
$x_m$ is the value of $x$ at which the maximum of the distribution occurs,
$b$ is the asymmetry parameter, and $D$ is related to the full-width half-maximum (FWHM) of 
the distribution, $W$, by
\begin{equation}
\label{13} 
   W=D\frac{\sinh(b)}{b} .
\end{equation}
The green-dotted curve in Fig. \ref{fig8}a is a fit of the black-solid distribution with
$b=0.37$ and $D=0.0116$, while $pdf_m$ and $x_m$ are taken from the calculated
distribution. That fit, and the corresponding one for the black-solid curve in Fig. \ref{fig8}b
give, respectively, $W_{nl} \simeq 0.012$ and $W_{loc} \simeq 0.0008$ for the non-locally 
and locally coupled SQUID metamaterial.  For the quantification of the metastability level, 
we use here the FWHM of the distributions, which for a symmetric Gaussian is 
directly proportional to the standard deviation $\sigma_{|\Psi|}$. Similarly, the squares 
of $W_{nl}$ and  $W_{loc}$ are proportional to the variance $\sigma_{|\Psi|}^2$ which is 
a measure of the metastability level. The squares of the calculated FWHMs, $W_{nl}$ and 
$W_{loc}$, differ by more than two orders of magnitude, i.e., $(W_{nl}/W_{loc})^2 \simeq 225$,
indicating the high metastability level of the chimera state compared to that of the 
corresponding non-uniform state. 
Note that for uniform states, that result for zero initial conditions both for non-locally 
and locally coupled SQUID metamaterials, the FWHM is pactically zero.
In Fig. \ref{fig4}b, the high metastability level results in breaking the quasi-stationarity 
of the chimera state that lasts for $\sim 0.35\times 10^7$ t.u. through sudden expansions
of the upper asynchronous cluster. In that case, the corresponding distribution looks like 
a "multimodal" one with three peaks (red curve of Fig. \ref{fig8}a); each of them may be 
however fitted to a skewed Gaussian of the form of Eq. (\ref{12}). 
The FWHM for each peak is roughly two order of magnitudes larger than that of the distribution
of the corresponding non-uniform state (red curve in Fig. \ref{fig8}b).

\section{Conclusions}
Chimera states are surprising spatiotemporal patterns in which regions of coherence and 
incoherence coexist. While they were initially discovered in numerical simulations, they 
have been subsequently observed in several experiments. Notably, these experiments do not 
merely confirm the numerical and theoretical predictions, but they also reveal other types 
of interesting collective dynamic behaviour. SQUID metamaterials, which comprise non-locally
coupled, nonlinear resonant elements, are physical systems where chimeras or other collective 
effects could be in principle detected. The emergence of long-lived chimera states in SQUID
metamaterials could be acheived by randomly initializing the fluxes threading the SQUID 
rings. Random initial flux configurations could be achieved by thermal quenching, i.e., 
by cooling down the SQUID metamaterial through its superconducting transition temperature.
As it has been demonstrated in superconducting loops \cite{Monaco2009} and superconducting
thin-film rings \cite{Kirtley2003}, spontaneous fluxoid formation may occur through the 
Kibble-Zurek scenario or by thermal activation, respectively. The Kibble-Zurek scenario
has been also confirmed in a multi-Josephson junction superconducting loop \cite{Carmi2000}.
In SQUIDs, however, fluxoid conservation is only approximate, and practically non-existent
in non-hysteretic SQUIDs that are considered here. Thus, we would expect that thermal 
quenching of the SQUID metamaterial could result in a random initial flux 
configuration, in which the fluxes can take any value (i.e., not only integer multiples
of the flux quantum). Note that the random flux configuration is just one possible 
choice of initial conditions for which chimera states could be observed in SQUID metamaterials.

Our numerical simulations rely on a realistic model,
which is capable of reproducing experimental findings such as the tuneability patterns of 
two-dimensional SQUID metamaterials which are obtained by varying an applied dc flux
\cite{Trepanier2013,Tsironis2014b}. These simulations indicate that in many cases the coherent
and incoherent clusters that make a chimera state maintain their individual sizes for very
long times ($>10^7$ t.u.). During the integration, induced instabilities may lead to sudden 
expansion of the incoherent cluster(s), or to the desynchronization of narrow coherent 
clusters. The inherent metastability of the chimera states presented here for the non-locally 
coupled SQUID metamaterials can be quantified by the strength of the fluctuations of
$<|\Psi (\tau)|>_T$. The corresponding locally coupled SQUID metamaterials, in which the 
interaction between SQUIDs is limited to their first neighbors, supports non-uniform solutions 
which however are not chimeras. 
Although both types of states may exhibit a low degree of synchronization, monitored as a
function of $\tau$ by $<|\Psi (\tau)|>_T$, the fluctuations of the latter in time are much
stronger for chimera states than for non-uniform states (more than two orders of magnitude,
for the cases shown). Due to their high metastability levels, chimera states are very 
sensitive to even slight parameter variations.
SQUID metamaterials, both in one and two dimensions, have been already fabricated, and 
some of their dynamical properties have been investigated \cite{Jung2014}. Thus, chimera
states could in principle be detected with the presently available experimental set-ups.

\section*{Acknowledgements}
This work was partially supported by
the European Union Seventh Framework Programme (FP7-REGPOT-2012-2013-1)
under grant agreement n$^o$ 316165, and
by the Thales Project MACOMSYS, co‐financed by the European Union
(European Social Fund – ESF) and Greek national funds through the Operational
Program "Education and Lifelong Learning" of the National Strategic Reference
Framework (NSRF) ‐ Research Funding Program: THALES
Investing in Knowledge Society via the European Social Fund.
NL and GPT thank Prof. Steven Anlage for illuminating discussions.


\begin{thebibliography}{58}
\expandafter\ifx\csname natexlab\endcsname\relax\def\natexlab#1{#1}\fi
\expandafter\ifx\csname bibnamefont\endcsname\relax
  \def\bibnamefont#1{#1}\fi
\expandafter\ifx\csname bibfnamefont\endcsname\relax
  \def\bibfnamefont#1{#1}\fi
\expandafter\ifx\csname citenamefont\endcsname\relax
  \def\citenamefont#1{#1}\fi
\expandafter\ifx\csname url\endcsname\relax
  \def\url#1{\texttt{#1}}\fi
\expandafter\ifx\csname urlprefix\endcsname\relax\def\urlprefix{URL }\fi
\providecommand{\bibinfo}[2]{#2}
\providecommand{\eprint}[2][]{\url{#2}}

\bibitem[{\citenamefont{Anlage}(2011)}]{Anlage2011}
\bibinfo{author}{\bibfnamefont{S.~M.} \bibnamefont{Anlage}},
  \bibinfo{journal}{J. Opt.} \textbf{\bibinfo{volume}{13}},
  \bibinfo{pages}{024001} (\bibinfo{year}{2011}).

\bibitem[{\citenamefont{Jung et~al.}(2014{\natexlab{a}})\citenamefont{Jung,
  Ustinov, and Anlage}}]{Jung2014}
\bibinfo{author}{\bibfnamefont{P.}~\bibnamefont{Jung}},
  \bibinfo{author}{\bibfnamefont{A.~V.} \bibnamefont{Ustinov}},
  \bibnamefont{and} \bibinfo{author}{\bibfnamefont{S.~M.}
  \bibnamefont{Anlage}}, \bibinfo{journal}{Supercond. Sci. Technol.}
  \textbf{\bibinfo{volume}{27}}, \bibinfo{pages}{073001}
  (\bibinfo{year}{2014}{\natexlab{a}}).

\bibitem[{\citenamefont{Jung et~al.}(2013)\citenamefont{Jung, Butz, Shitov, and
  Ustinov}}]{Jung2013}
\bibinfo{author}{\bibfnamefont{P.}~\bibnamefont{Jung}},
  \bibinfo{author}{\bibfnamefont{S.}~\bibnamefont{Butz}},
  \bibinfo{author}{\bibfnamefont{S.~V.} \bibnamefont{Shitov}},
  \bibnamefont{and} \bibinfo{author}{\bibfnamefont{A.~V.}
  \bibnamefont{Ustinov}}, \bibinfo{journal}{Appl. Phys. Lett.}
  \textbf{\bibinfo{volume}{102}}, \bibinfo{pages}{062601}
  (\bibinfo{year}{2013}).

\bibitem[{\citenamefont{Butz et~al.}(2013)\citenamefont{Butz, Jung, Filippenko,
  Koshelets, and Ustinov}}]{Butz2013a}
\bibinfo{author}{\bibfnamefont{S.}~\bibnamefont{Butz}},
  \bibinfo{author}{\bibfnamefont{P.}~\bibnamefont{Jung}},
  \bibinfo{author}{\bibfnamefont{L.~V.} \bibnamefont{Filippenko}},
  \bibinfo{author}{\bibfnamefont{V.~P.} \bibnamefont{Koshelets}},
  \bibnamefont{and} \bibinfo{author}{\bibfnamefont{A.~V.}
  \bibnamefont{Ustinov}}, \bibinfo{journal}{Opt. Express}
  \textbf{\bibinfo{volume}{29}}, \bibinfo{pages}{22540}
  (\bibinfo{year}{2013}).

\bibitem[{\citenamefont{Trepanier et~al.}(2013)\citenamefont{Trepanier, Zhang,
  Mukhanov, and Anlage}}]{Trepanier2013}
\bibinfo{author}{\bibfnamefont{M.}~\bibnamefont{Trepanier}},
  \bibinfo{author}{\bibfnamefont{D.}~\bibnamefont{Zhang}},
  \bibinfo{author}{\bibfnamefont{O.}~\bibnamefont{Mukhanov}}, \bibnamefont{and}
  \bibinfo{author}{\bibfnamefont{S.~M.} \bibnamefont{Anlage}},
  \bibinfo{journal}{Phys. Rev. X} \textbf{\bibinfo{volume}{3}},
  \bibinfo{pages}{041029} (\bibinfo{year}{2013}).

\bibitem[{\citenamefont{Jung et~al.}(2014{\natexlab{b}})\citenamefont{Jung,
  Butz, Marthaler, Fistul, Lepp{\"a}kangas, Koshelets, and
  Ustinov}}]{Jung2014b}
\bibinfo{author}{\bibfnamefont{P.}~\bibnamefont{Jung}},
  \bibinfo{author}{\bibfnamefont{S.}~\bibnamefont{Butz}},
  \bibinfo{author}{\bibfnamefont{M.}~\bibnamefont{Marthaler}},
  \bibinfo{author}{\bibfnamefont{M.~V.} \bibnamefont{Fistul}},
  \bibinfo{author}{\bibfnamefont{J.}~\bibnamefont{Lepp{\"a}kangas}},
  \bibinfo{author}{\bibfnamefont{V.~P.} \bibnamefont{Koshelets}},
  \bibnamefont{and} \bibinfo{author}{\bibfnamefont{A.~V.}
  \bibnamefont{Ustinov}}, \bibinfo{journal}{Nat. Comms.}
  \textbf{\bibinfo{volume}{5}}, \bibinfo{pages}{3730}
  (\bibinfo{year}{2014}{\natexlab{b}}).

\bibitem[{\citenamefont{Josephson}(1962)}]{Josephson1962}
\bibinfo{author}{\bibfnamefont{B.}~\bibnamefont{Josephson}},
  \bibinfo{journal}{Phys. Lett. A} \textbf{\bibinfo{volume}{1}},
  \bibinfo{pages}{251} (\bibinfo{year}{1962}).

\bibitem[{\citenamefont{Du et~al.}(2006)\citenamefont{Du, Chen, and
  Li}}]{Du2006}
\bibinfo{author}{\bibfnamefont{C.}~\bibnamefont{Du}},
  \bibinfo{author}{\bibfnamefont{H.}~\bibnamefont{Chen}}, \bibnamefont{and}
  \bibinfo{author}{\bibfnamefont{S.}~\bibnamefont{Li}}, \bibinfo{journal}{Phys.
  Rev. B} \textbf{\bibinfo{volume}{74}}, \bibinfo{pages}{113105}
  (\bibinfo{year}{2006}).

\bibitem[{\citenamefont{Lazarides and Tsironis}(2007)}]{Lazarides2007}
\bibinfo{author}{\bibfnamefont{N.}~\bibnamefont{Lazarides}} \bibnamefont{and}
  \bibinfo{author}{\bibfnamefont{G.~P.} \bibnamefont{Tsironis}},
  \bibinfo{journal}{Appl. Phys. Lett.} \textbf{\bibinfo{volume}{16}},
  \bibinfo{pages}{163501} (\bibinfo{year}{2007}).

\bibitem[{\citenamefont{Strogatz}(2001)}]{Strogatz2001}
\bibinfo{author}{\bibfnamefont{S.~H.} \bibnamefont{Strogatz}},
  \bibinfo{journal}{Nature} \textbf{\bibinfo{volume}{410}},
  \bibinfo{pages}{268} (\bibinfo{year}{2001}).

\bibitem[{\citenamefont{Strogatz}(2000)}]{Strogatz2000}
\bibinfo{author}{\bibfnamefont{S.~H.} \bibnamefont{Strogatz}},
  \bibinfo{journal}{Physica D} \textbf{\bibinfo{volume}{143}},
  \bibinfo{pages}{1} (\bibinfo{year}{2000}).

\bibitem[{\citenamefont{Acebr{\'o}n et~al.}(2005)\citenamefont{Acebr{\'o}n,
  Bonilla, Vicente, Ritort, and Spigler}}]{Acebron2005}
\bibinfo{author}{\bibfnamefont{J.~A.} \bibnamefont{Acebr{\'o}n}},
  \bibinfo{author}{\bibfnamefont{L.~L.} \bibnamefont{Bonilla}},
  \bibinfo{author}{\bibfnamefont{C.~J.~P.} \bibnamefont{Vicente}},
  \bibinfo{author}{\bibfnamefont{F.}~\bibnamefont{Ritort}}, \bibnamefont{and}
  \bibinfo{author}{\bibfnamefont{R.}~\bibnamefont{Spigler}},
  \bibinfo{journal}{Rev. Mod. Phys.} \textbf{\bibinfo{volume}{77}},
  \bibinfo{pages}{135} (\bibinfo{year}{2005}).

\bibitem[{\citenamefont{Battogtokh}(1999)}]{Battogtokh1999}
\bibinfo{author}{\bibfnamefont{D.}~\bibnamefont{Battogtokh}},
  \bibinfo{journal}{Prog. Theor. Phys.} \textbf{\bibinfo{volume}{102}},
  \bibinfo{pages}{947} (\bibinfo{year}{1999}).

\bibitem[{\citenamefont{Viana et~al.}(2011)\citenamefont{Viana, dos S.~Silva,
  and Lopes}}]{Viana2011}
\bibinfo{author}{\bibfnamefont{R.~L.} \bibnamefont{Viana}},
  \bibinfo{author}{\bibfnamefont{F.~A.} \bibnamefont{dos S.~Silva}},
  \bibnamefont{and} \bibinfo{author}{\bibfnamefont{S.~R.} \bibnamefont{Lopes}},
  \bibinfo{journal}{Phys. Rev. E} \textbf{\bibinfo{volume}{83}},
  \bibinfo{pages}{046220} (\bibinfo{year}{2011}).

\bibitem[{\citenamefont{Kuramoto and Battogtokh}(2002)}]{Kuramoto2002}
\bibinfo{author}{\bibfnamefont{Y.}~\bibnamefont{Kuramoto}} \bibnamefont{and}
  \bibinfo{author}{\bibfnamefont{D.}~\bibnamefont{Battogtokh}},
  \bibinfo{journal}{Nonlinear Phenom. Complex Syst.}
  \textbf{\bibinfo{volume}{5}}, \bibinfo{pages}{380} (\bibinfo{year}{2002}).

\bibitem[{\citenamefont{Abrams and Strogatz}(2004)}]{Abrams2004}
\bibinfo{author}{\bibfnamefont{D.~M.} \bibnamefont{Abrams}} \bibnamefont{and}
  \bibinfo{author}{\bibfnamefont{S.~H.} \bibnamefont{Strogatz}},
  \bibinfo{journal}{Phys. Rev. Lett.} \textbf{\bibinfo{volume}{93}},
  \bibinfo{pages}{174102} (\bibinfo{year}{2004}).

\bibitem[{\citenamefont{Kuramoto et~al.}(2006)\citenamefont{Kuramoto, Shima,
  Battogtokh, and Shiogai}}]{Kuramoto2006}
\bibinfo{author}{\bibfnamefont{Y.}~\bibnamefont{Kuramoto}},
  \bibinfo{author}{\bibfnamefont{S.-I.} \bibnamefont{Shima}},
  \bibinfo{author}{\bibfnamefont{D.}~\bibnamefont{Battogtokh}},
  \bibnamefont{and} \bibinfo{author}{\bibfnamefont{Y.}~\bibnamefont{Shiogai}},
  \bibinfo{journal}{Prog. Theor. Phys., Suppl.} \textbf{\bibinfo{volume}{161}},
  \bibinfo{pages}{127} (\bibinfo{year}{2006}).

\bibitem[{\citenamefont{Omel'chenko et~al.}(2008)\citenamefont{Omel'chenko,
  Maistrenko, and Tass}}]{Omelchenko2008}
\bibinfo{author}{\bibfnamefont{O.~E.} \bibnamefont{Omel'chenko}},
  \bibinfo{author}{\bibfnamefont{Y.~L.} \bibnamefont{Maistrenko}},
  \bibnamefont{and} \bibinfo{author}{\bibfnamefont{P.~A.} \bibnamefont{Tass}},
  \bibinfo{journal}{Phys. Rev. Lett.} \textbf{\bibinfo{volume}{100}},
  \bibinfo{pages}{044105} (\bibinfo{year}{2008}).

\bibitem[{\citenamefont{Abrams et~al.}(2008)\citenamefont{Abrams, Mirollo,
  Strogatz, and Wiley}}]{Abrams2008}
\bibinfo{author}{\bibfnamefont{D.~M.} \bibnamefont{Abrams}},
  \bibinfo{author}{\bibfnamefont{R.}~\bibnamefont{Mirollo}},
  \bibinfo{author}{\bibfnamefont{S.~H.} \bibnamefont{Strogatz}},
  \bibnamefont{and} \bibinfo{author}{\bibfnamefont{D.~A.} \bibnamefont{Wiley}},
  \bibinfo{journal}{Phys. Rev. Lett.} \textbf{\bibinfo{volume}{101}},
  \bibinfo{pages}{084103} (\bibinfo{year}{2008}).

\bibitem[{\citenamefont{Pikovsky and Rosenblum}(2008)}]{Pikovsky2008b}
\bibinfo{author}{\bibfnamefont{A.}~\bibnamefont{Pikovsky}} \bibnamefont{and}
  \bibinfo{author}{\bibfnamefont{M.}~\bibnamefont{Rosenblum}},
  \bibinfo{journal}{Phys. Rev. Lett.} \textbf{\bibinfo{volume}{101}},
  \bibinfo{pages}{264103} (\bibinfo{year}{2008}).

\bibitem[{\citenamefont{Ott and Antonsen}(2009)}]{Ott2009}
\bibinfo{author}{\bibfnamefont{E.}~\bibnamefont{Ott}} \bibnamefont{and}
  \bibinfo{author}{\bibfnamefont{T.~M.} \bibnamefont{Antonsen}},
  \bibinfo{journal}{Chaos} \textbf{\bibinfo{volume}{19}},
  \bibinfo{pages}{023117} (\bibinfo{year}{2009}).

\bibitem[{\citenamefont{Martens et~al.}(2010)\citenamefont{Martens, Laing, and
  Strogatz}}]{Martens2010}
\bibinfo{author}{\bibfnamefont{E.~A.} \bibnamefont{Martens}},
  \bibinfo{author}{\bibfnamefont{C.~R.} \bibnamefont{Laing}}, \bibnamefont{and}
  \bibinfo{author}{\bibfnamefont{S.~H.} \bibnamefont{Strogatz}},
  \bibinfo{journal}{Phys. Rev. Lett.} \textbf{\bibinfo{volume}{104}},
  \bibinfo{pages}{044101} (\bibinfo{year}{2010}).

\bibitem[{\citenamefont{Omelchenko et~al.}(2011)\citenamefont{Omelchenko,
  Maistrenko, H{\"o}vel, and Sch{\"o}ll}}]{Omelchenko2011}
\bibinfo{author}{\bibfnamefont{I.}~\bibnamefont{Omelchenko}},
  \bibinfo{author}{\bibfnamefont{Y.}~\bibnamefont{Maistrenko}},
  \bibinfo{author}{\bibfnamefont{P.}~\bibnamefont{H{\"o}vel}},
  \bibnamefont{and}
  \bibinfo{author}{\bibfnamefont{E.}~\bibnamefont{Sch{\"o}ll}},
  \bibinfo{journal}{Phys. Rev. Lett.} \textbf{\bibinfo{volume}{106}},
  \bibinfo{pages}{234102} (\bibinfo{year}{2011}).

\bibitem[{\citenamefont{Yao et~al.}(2013)\citenamefont{Yao, Huang, Lai, and
  Zheng}}]{Yao2013}
\bibinfo{author}{\bibfnamefont{N.}~\bibnamefont{Yao}},
  \bibinfo{author}{\bibfnamefont{Z.-G.} \bibnamefont{Huang}},
  \bibinfo{author}{\bibfnamefont{Y.-C.} \bibnamefont{Lai}}, \bibnamefont{and}
  \bibinfo{author}{\bibfnamefont{Z.-G.} \bibnamefont{Zheng}},
  \bibinfo{journal}{Sci. Rep.} \textbf{\bibinfo{volume}{3}},
  \bibinfo{pages}{3522} (\bibinfo{year}{2013}).

\bibitem[{\citenamefont{Omelchenko et~al.}(2013)\citenamefont{Omelchenko,
  Omel'chenko, H{\"o}vel, and Sch{\"o}ll}}]{Omelchenko2013}
\bibinfo{author}{\bibfnamefont{I.}~\bibnamefont{Omelchenko}},
  \bibinfo{author}{\bibfnamefont{O.~E.} \bibnamefont{Omel'chenko}},
  \bibinfo{author}{\bibfnamefont{P.}~\bibnamefont{H{\"o}vel}},
  \bibnamefont{and}
  \bibinfo{author}{\bibfnamefont{E.}~\bibnamefont{Sch{\"o}ll}},
  \bibinfo{journal}{Phys. Rev. Lett.} \textbf{\bibinfo{volume}{110}},
  \bibinfo{pages}{224101} (\bibinfo{year}{2013}).

\bibitem[{\citenamefont{Hizanidis et~al.}(2014)\citenamefont{Hizanidis, Kanas,
  Bezerianos, and Bountis}}]{Hizanidis2014}
\bibinfo{author}{\bibfnamefont{J.}~\bibnamefont{Hizanidis}},
  \bibinfo{author}{\bibfnamefont{V.}~\bibnamefont{Kanas}},
  \bibinfo{author}{\bibfnamefont{A.}~\bibnamefont{Bezerianos}},
  \bibnamefont{and} \bibinfo{author}{\bibfnamefont{T.}~\bibnamefont{Bountis}},
  \bibinfo{journal}{Int. J. Bifurcation Chaos} \textbf{\bibinfo{volume}{24
  }}, \bibinfo{pages}{1450030} (\bibinfo{year}{2014}).

\bibitem[{\citenamefont{Zakharova et~al.}(2014)\citenamefont{Zakharova,
  Kapeller, and Sch{\"o}ll}}]{Zakharova2014}
\bibinfo{author}{\bibfnamefont{A.}~\bibnamefont{Zakharova}},
  \bibinfo{author}{\bibfnamefont{M.}~\bibnamefont{Kapeller}}, \bibnamefont{and}
  \bibinfo{author}{\bibfnamefont{E.}~\bibnamefont{Sch{\"o}ll}},
  \bibinfo{journal}{Phys. Rev. Lett.} \textbf{\bibinfo{volume}{112}},
  \bibinfo{pages}{154101} (\bibinfo{year}{2014}).

\bibitem[{\citenamefont{Bountis et~al.}(2014)\citenamefont{Bountis, Kanas,
  Hizanidis, and Bezerianos}}]{Bountis2014}
\bibinfo{author}{\bibfnamefont{T.}~\bibnamefont{Bountis}},
  \bibinfo{author}{\bibfnamefont{V.}~\bibnamefont{Kanas}},
  \bibinfo{author}{\bibfnamefont{J.}~\bibnamefont{Hizanidis}},
  \bibnamefont{and}
  \bibinfo{author}{\bibfnamefont{A.}~\bibnamefont{Bezerianos}},
  \bibinfo{journal}{Eur. Phys. J.-Spec. Top.} \textbf{\bibinfo{volume}{223}},
  \bibinfo{pages}{721} (\bibinfo{year}{2014}).

\bibitem[{\citenamefont{Yeldesbay et~al.}(2014)\citenamefont{Yeldesbay,
  Pikovsky, and Rosenblum}}]{Yeldesbay2014}
\bibinfo{author}{\bibfnamefont{A.}~\bibnamefont{Yeldesbay}},
  \bibinfo{author}{\bibfnamefont{A.}~\bibnamefont{Pikovsky}}, \bibnamefont{and}
  \bibinfo{author}{\bibfnamefont{M.}~\bibnamefont{Rosenblum}},
  \bibinfo{journal}{Phys. Rev. Lett.} \textbf{\bibinfo{volume}{112}},
  \bibinfo{pages}{144103} (\bibinfo{year}{2014}).

\bibitem[{\citenamefont{Tinsley et~al.}(2012)\citenamefont{Tinsley, Nkomo, and
  Showalter}}]{Tinsley2012}
\bibinfo{author}{\bibfnamefont{M.~R.} \bibnamefont{Tinsley}},
  \bibinfo{author}{\bibfnamefont{S.}~\bibnamefont{Nkomo}}, \bibnamefont{and}
  \bibinfo{author}{\bibfnamefont{K.}~\bibnamefont{Showalter}},
  \bibinfo{journal}{Nature Phys.} \textbf{\bibinfo{volume}{8}},
  \bibinfo{pages}{662} (\bibinfo{year}{2012}).

\bibitem[{\citenamefont{Hagerstrom et~al.}(2012)\citenamefont{Hagerstrom,
  Murphy, Roy, H{\"o}vel, Omelchenko, and Sch{\"o}ll}}]{Hagerstrom2012}
\bibinfo{author}{\bibfnamefont{A.~M.} \bibnamefont{Hagerstrom}},
  \bibinfo{author}{\bibfnamefont{T.~E.} \bibnamefont{Murphy}},
  \bibinfo{author}{\bibfnamefont{R.}~\bibnamefont{Roy}},
  \bibinfo{author}{\bibfnamefont{P.}~\bibnamefont{H{\"o}vel}},
  \bibinfo{author}{\bibfnamefont{I.}~\bibnamefont{Omelchenko}},
  \bibnamefont{and}
  \bibinfo{author}{\bibfnamefont{E.}~\bibnamefont{Sch{\"o}ll}},
  \bibinfo{journal}{Nature Phys.} \textbf{\bibinfo{volume}{8}},
  \bibinfo{pages}{658} (\bibinfo{year}{2012}).

\bibitem[{\citenamefont{Wickramasinghe and Kiss}(2013)}]{Wickra2013}
\bibinfo{author}{\bibfnamefont{M.}~\bibnamefont{Wickramasinghe}}
  \bibnamefont{and} \bibinfo{author}{\bibfnamefont{I.~Z.} \bibnamefont{Kiss}},
  \bibinfo{journal}{PLOS ONE} \textbf{\bibinfo{volume}{8}},
  \bibinfo{pages}{e80586 [12 pages]} (\bibinfo{year}{2013}).

\bibitem[{\citenamefont{Nkomo et~al.}(2013)\citenamefont{Nkomo, Tinsley, and
  Showalter}}]{Nkomo2013}
\bibinfo{author}{\bibfnamefont{S.}~\bibnamefont{Nkomo}},
  \bibinfo{author}{\bibfnamefont{M.~R.} \bibnamefont{Tinsley}},
  \bibnamefont{and}
  \bibinfo{author}{\bibfnamefont{K.}~\bibnamefont{Showalter}},
  \bibinfo{journal}{Phys. Rev. Lett.} \textbf{\bibinfo{volume}{110}},
  \bibinfo{pages}{244102} (\bibinfo{year}{2013}).

\bibitem[{\citenamefont{Martens et~al.}(2013)\citenamefont{Martens, Thutupalli,
  Fourri{\'e}re, and Hallatschek}}]{Martens2013}
\bibinfo{author}{\bibfnamefont{E.~A.} \bibnamefont{Martens}},
  \bibinfo{author}{\bibfnamefont{S.}~\bibnamefont{Thutupalli}},
  \bibinfo{author}{\bibfnamefont{A.}~\bibnamefont{Fourri{\'e}re}},
  \bibnamefont{and}
  \bibinfo{author}{\bibfnamefont{O.}~\bibnamefont{Hallatschek}},
  \bibinfo{journal}{Proc. Natl. Acad. Sci.} \textbf{\bibinfo{volume}{110}},
  \bibinfo{pages}{10563} (\bibinfo{year}{2013}).

\bibitem[{\citenamefont{Sch{\"o}nleber
  et~al.}(2014)\citenamefont{Sch{\"o}nleber, Zensen, Heinrich, and
  Krischer}}]{Schonleber2014}
\bibinfo{author}{\bibfnamefont{K.}~\bibnamefont{Sch{\"o}nleber}},
  \bibinfo{author}{\bibfnamefont{C.}~\bibnamefont{Zensen}},
  \bibinfo{author}{\bibfnamefont{A.}~\bibnamefont{Heinrich}}, \bibnamefont{and}
  \bibinfo{author}{\bibfnamefont{K.}~\bibnamefont{Krischer}},
  \bibinfo{journal}{New J. Phys.} \textbf{\bibinfo{volume}{16}},
  \bibinfo{pages}{063024} (\bibinfo{year}{2014}).

\bibitem[{\citenamefont{Viktorov et~al.}(2014)\citenamefont{Viktorov,
  Habruseva, Hegarty, Huyet, and Kelleher}}]{Viktorov2014}
\bibinfo{author}{\bibfnamefont{E.~A.} \bibnamefont{Viktorov}},
  \bibinfo{author}{\bibfnamefont{T.}~\bibnamefont{Habruseva}},
  \bibinfo{author}{\bibfnamefont{S.~P.} \bibnamefont{Hegarty}},
  \bibinfo{author}{\bibfnamefont{G.}~\bibnamefont{Huyet}}, \bibnamefont{and}
  \bibinfo{author}{\bibfnamefont{B.}~\bibnamefont{Kelleher}},
  \bibinfo{journal}{Phys. Rev. Lett.} \textbf{\bibinfo{volume}{112}},
  \bibinfo{pages}{224101} (\bibinfo{year}{2014}).

\bibitem[{\citenamefont{Rosin et~al.}(2014)\citenamefont{Rosin, Rontani,
  Haynes, Sch{\"u}ll, and Gauthier}}]{Rosin2014}
\bibinfo{author}{\bibfnamefont{D.~P.} \bibnamefont{Rosin}},
  \bibinfo{author}{\bibfnamefont{D.}~\bibnamefont{Rontani}},
  \bibinfo{author}{\bibfnamefont{N.~D.} \bibnamefont{Haynes}},
  \bibinfo{author}{\bibfnamefont{E.}~\bibnamefont{Sch{\"u}ll}},
  \bibnamefont{and} \bibinfo{author}{\bibfnamefont{D.~J.}
  \bibnamefont{Gauthier}}, \bibinfo{journal}{arXiv:}
  \textbf{\bibinfo{volume}{1405.1950}} (\bibinfo{year}{2014}).

\bibitem[{\citenamefont{Schmidt et~al.}(2014)\citenamefont{Schmidt,
  Sch{\"o}nleber, Krischer, and Garc{\'i}a-Morales}}]{Schmidt2014b}
\bibinfo{author}{\bibfnamefont{L.}~\bibnamefont{Schmidt}},
  \bibinfo{author}{\bibfnamefont{K.}~\bibnamefont{Sch{\"o}nleber}},
  \bibinfo{author}{\bibfnamefont{K.}~\bibnamefont{Krischer}}, \bibnamefont{and}
  \bibinfo{author}{\bibfnamefont{V.}~\bibnamefont{Garc{\'i}a-Morales}},
  \bibinfo{journal}{Chaos} \textbf{\bibinfo{volume}{24}},
  \bibinfo{pages}{013102} (\bibinfo{year}{2014}).

\bibitem[{\citenamefont{Gambuzza et~al.}(2014)\citenamefont{Gambuzza,
  Buscarino, Chessari, Fortuna, Meucci, and Frasca}}]{Gambuzza2014}
\bibinfo{author}{\bibfnamefont{L.~V.} \bibnamefont{Gambuzza}},
  \bibinfo{author}{\bibfnamefont{A.}~\bibnamefont{Buscarino}},
  \bibinfo{author}{\bibfnamefont{S.}~\bibnamefont{Chessari}},
  \bibinfo{author}{\bibfnamefont{L.}~\bibnamefont{Fortuna}},
  \bibinfo{author}{\bibfnamefont{R.}~\bibnamefont{Meucci}}, \bibnamefont{and}
  \bibinfo{author}{\bibfnamefont{M.}~\bibnamefont{Frasca}},
  \bibinfo{journal}{Phys. Rev. E} \textbf{\bibinfo{volume}{90}},
  \bibinfo{pages}{032905} (\bibinfo{year}{2014}).

\bibitem[{\citenamefont{Kapitaniak et~al.}(2014)\citenamefont{Kapitaniak,
  Kuzma, Wojewoda, Czolczynski, and Maistrenko}}]{Kapitaniak2014}
\bibinfo{author}{\bibfnamefont{T.}~\bibnamefont{Kapitaniak}},
  \bibinfo{author}{\bibfnamefont{P.}~\bibnamefont{Kuzma}},
  \bibinfo{author}{\bibfnamefont{J.}~\bibnamefont{Wojewoda}},
  \bibinfo{author}{\bibfnamefont{K.}~\bibnamefont{Czolczynski}},
  \bibnamefont{and}
  \bibinfo{author}{\bibfnamefont{Y.}~\bibnamefont{Maistrenko}},
  \bibinfo{journal}{Sci. Rep.} \textbf{\bibinfo{volume}{4}},
  \bibinfo{pages}{6379} (\bibinfo{year}{2014}).

\bibitem[{\citenamefont{Smart}(2012)}]{Smart2012}
\bibinfo{author}{\bibfnamefont{A.~G.} \bibnamefont{Smart}},
  \bibinfo{journal}{Physics Today} \textbf{\bibinfo{volume}{65}},
  \bibinfo{pages}{17} (\bibinfo{year}{2012}).

\bibitem[{\citenamefont{Panaggio and Abrams}(2014)}]{Panaggio2014}
\bibinfo{author}{\bibfnamefont{M.~J.} \bibnamefont{Panaggio}} \bibnamefont{and}
  \bibinfo{author}{\bibfnamefont{D.~M.} \bibnamefont{Abrams}},
  \bibinfo{journal}{arXiv:} \textbf{\bibinfo{volume}{1403.6204}}

\bibitem[{\citenamefont{Lazarides et~al.}(2008)\citenamefont{Lazarides,
  Tsironis, and Eleftheriou}}]{Lazarides2008a}
\bibinfo{author}{\bibfnamefont{N.}~\bibnamefont{Lazarides}},
  \bibinfo{author}{\bibfnamefont{G.~P.} \bibnamefont{Tsironis}},
  \bibnamefont{and}
  \bibinfo{author}{\bibfnamefont{M.}~\bibnamefont{Eleftheriou}},
  \bibinfo{journal}{Nonlinear Phenom. Complex Syst.}
  \textbf{\bibinfo{volume}{11}}, \bibinfo{pages}{250} (\bibinfo{year}{2008}).

\bibitem[{\citenamefont{Lazarides and Tsironis}(2012)}]{Lazarides2012}
\bibinfo{author}{\bibfnamefont{N.}~\bibnamefont{Lazarides}} \bibnamefont{and}
  \bibinfo{author}{\bibfnamefont{G.~P.} \bibnamefont{Tsironis}},
  \bibinfo{journal}{Proc. SPIE} \textbf{\bibinfo{volume}{8423}},
  \bibinfo{pages}{84231K} (\bibinfo{year}{2012}).

\bibitem[{\citenamefont{Lazarides and Tsironis}(2013)}]{Lazarides2013b}
\bibinfo{author}{\bibfnamefont{N.}~\bibnamefont{Lazarides}} \bibnamefont{and}
  \bibinfo{author}{\bibfnamefont{G.~P.} \bibnamefont{Tsironis}},
  \bibinfo{journal}{Supercond. Sci. Technol.} \textbf{\bibinfo{volume}{26}},
  \bibinfo{pages}{084006} (\bibinfo{year}{2013}).

\bibitem[{\citenamefont{Shanahan}(2010)}]{Shanahan2010}
\bibinfo{author}{\bibfnamefont{M.}~\bibnamefont{Shanahan}},
  \bibinfo{journal}{Chaos} \textbf{\bibinfo{volume}{20}},
  \bibinfo{pages}{013108} (\bibinfo{year}{2010}).

\bibitem[{\citenamefont{Wildie and Shanahan}(2012)}]{Wildie2012}
\bibinfo{author}{\bibfnamefont{M.}~\bibnamefont{Wildie}} \bibnamefont{and}
  \bibinfo{author}{\bibfnamefont{M.}~\bibnamefont{Shanahan}},
  \bibinfo{journal}{Chaos} \textbf{\bibinfo{volume}{22}},
  \bibinfo{pages}{043131} (\bibinfo{year}{2012}).

\bibitem[{\citenamefont{Wolfrum and Omel'chenko}(2011)}]{Wolfrum2011}
\bibinfo{author}{\bibfnamefont{M.}~\bibnamefont{Wolfrum}} \bibnamefont{and}
  \bibinfo{author}{\bibfnamefont{O.~E.} \bibnamefont{Omel'chenko}},
  \bibinfo{journal}{Phys. Rev. E} \textbf{\bibinfo{volume}{84}},
  \bibinfo{pages}{015201} (\bibinfo{year}{2011}).

\bibitem[{\citenamefont{Singh et~al.}(2011)\citenamefont{Singh, Dasgupta, and
  Sinha}}]{Singh2011}
\bibinfo{author}{\bibfnamefont{R.}~\bibnamefont{Singh}},
  \bibinfo{author}{\bibfnamefont{S.}~\bibnamefont{Dasgupta}}, \bibnamefont{and}
  \bibinfo{author}{\bibfnamefont{S.}~\bibnamefont{Sinha}},
  \bibinfo{journal}{Europhys. Lett.} \textbf{\bibinfo{volume}{95}},
  \bibinfo{pages}{10004} (\bibinfo{year}{2011}).

\bibitem[{\citenamefont{Sethia and Sen}(2014)}]{Sethia2014}
\bibinfo{author}{\bibfnamefont{G.~C.} \bibnamefont{Sethia}} \bibnamefont{and}
  \bibinfo{author}{\bibfnamefont{A.}~\bibnamefont{Sen}},
  \bibinfo{journal}{Phys. Rev. Lett.} \textbf{\bibinfo{volume}{112}},
  \bibinfo{pages}{144101} (\bibinfo{year}{2014}).

\bibitem[{\citenamefont{Campa et~al.}(2009)\citenamefont{Campa, Dauxois, and
  Ruffo}}]{Campa2009}
\bibinfo{author}{\bibfnamefont{A.}~\bibnamefont{Campa}},
  \bibinfo{author}{\bibfnamefont{T.}~\bibnamefont{Dauxois}}, \bibnamefont{and}
  \bibinfo{author}{\bibfnamefont{S.}~\bibnamefont{Ruffo}},
  \bibinfo{journal}{Phys. Rep.} \textbf{\bibinfo{volume}{480}},
  \bibinfo{pages}{57} (\bibinfo{year}{2009}).

\bibitem[{\citenamefont{Likharev.}(1986)}]{Likharev1986}
\bibinfo{author}{\bibfnamefont{K.~K.} \bibnamefont{Likharev.}},
  \emph{\bibinfo{title}{Dynamics of Josephson Junctions and Circuits.}}
  (\bibinfo{publisher}{Gordon and Breach}, \bibinfo{address}{Philadelphia},
  \bibinfo{year}{1986}).

\bibitem[{\citenamefont{Fraser and Suzuki}(1969)}]{Fraser1969}
\bibinfo{author}{\bibfnamefont{R.~D.~B.} \bibnamefont{Fraser}}
  \bibnamefont{and} \bibinfo{author}{\bibfnamefont{E.}~\bibnamefont{Suzuki}},
  \bibinfo{journal}{Anal. Chem.} \textbf{\bibinfo{volume}{41}},
  \bibinfo{pages}{37} (\bibinfo{year}{1969}).

\bibitem[{\citenamefont{Rusch and Lelieur}(1973)}]{Rusch1973}
\bibinfo{author}{\bibfnamefont{P.~F.} \bibnamefont{Rusch}} \bibnamefont{and}
  \bibinfo{author}{\bibfnamefont{J.~P.} \bibnamefont{Lelieur}},
  \bibinfo{journal}{Anal. Chem.} \textbf{\bibinfo{volume}{45}},
  \bibinfo{pages}{1541} (\bibinfo{year}{1973}).

\bibitem[{\citenamefont{Monaco et~al.}(2009)\citenamefont{Monaco, Mygind,
  Rivers, and Koshelets}}]{Monaco2009}
\bibinfo{author}{\bibfnamefont{R.}~\bibnamefont{Monaco}},
  \bibinfo{author}{\bibfnamefont{J.}~\bibnamefont{Mygind}},
  \bibinfo{author}{\bibfnamefont{R.~J.} \bibnamefont{Rivers}},
  \bibnamefont{and} \bibinfo{author}{\bibfnamefont{V.~P.}
  \bibnamefont{Koshelets}}, \bibinfo{journal}{Phys. Rev. B}
  \textbf{\bibinfo{volume}{80}}, \bibinfo{pages}{180501(R)}
  (\bibinfo{year}{2009}).

\bibitem[{\citenamefont{Kirtley et~al.}(2003)\citenamefont{Kirtley, Tsuei, and
  Tafuri}}]{Kirtley2003}
\bibinfo{author}{\bibfnamefont{J.~R.} \bibnamefont{Kirtley}},
  \bibinfo{author}{\bibfnamefont{C.~C.} \bibnamefont{Tsuei}}, \bibnamefont{and}
  \bibinfo{author}{\bibfnamefont{F.}~\bibnamefont{Tafuri}},
  \bibinfo{journal}{Phys. Rev. Lett.} \textbf{\bibinfo{volume}{90}},
  \bibinfo{pages}{257001} (\bibinfo{year}{2003}).

\bibitem[{\citenamefont{Carmi et~al.}(2000)\citenamefont{Carmi, Polturak, and
  Koren}}]{Carmi2000}
\bibinfo{author}{\bibfnamefont{R.}~\bibnamefont{Carmi}},
  \bibinfo{author}{\bibfnamefont{E.}~\bibnamefont{Polturak}}, \bibnamefont{and}
  \bibinfo{author}{\bibfnamefont{G.}~\bibnamefont{Koren}},
  \bibinfo{journal}{Phys. Rev. Lett.} \textbf{\bibinfo{volume}{84}},
  \bibinfo{pages}{4966} (\bibinfo{year}{2000}).

\bibitem[{\citenamefont{Tsironis et~al.}(2014)\citenamefont{Tsironis,
  Lazarides, and Margaris}}]{Tsironis2014b}
\bibinfo{author}{\bibfnamefont{G.~P.} \bibnamefont{Tsironis}},
  \bibinfo{author}{\bibfnamefont{N.}~\bibnamefont{Lazarides}},
  \bibnamefont{and} \bibinfo{author}{\bibfnamefont{I.}~\bibnamefont{Margaris}},
  \bibinfo{journal}{Appl. Phys. A} \textbf{\bibinfo{volume}{117}},
  \bibinfo{pages}{579} (\bibinfo{year}{2014}).

\end{thebibliography}


\end{document}